\documentclass[a4paper]{jpconf}
\usepackage{graphicx}
\usepackage{amssymb}
\begin{document}
\title{Leptogenesis with exclusively low-energy CP Violation in the Context of Minimal Lepton Flavour Violation}
\author{Selma Uhlig}
\address{Physik Department, Technische Universit\"at M\"unchen, D-85748 Garching, Germany}
\ead{selma.uhlig@ph.tum.de}
\begin{abstract}
We analyze lepton flavour violation (LFV) and the generation of the observed baryon-antibaryon asymmetry of the Universe (BAU) within a generalized minimal lepton flavour violation framework with three quasi-degenerate heavy Majorana neutrinos. The BAU which is obtained through radiative resonant leptogenesis can successfully be generated widely independent of the Majorana scale in this scenario and flavour effects are found to be relevant. Then we discuss the specific case in which CP violation is exclusively present at low-energies (a real $R$ matrix) in the flavour sensitive temperature regime. Successful leptogenesis in this case leads to strong constraints on low-energy neutrino parameters.
\end{abstract}
\section{Introduction}
Since the discovery of neutrino masses it is known that lepton flavour is not conserved.
However, from non-observation of LFV processes such as $\mu \to e
\gamma$ we know that those interactions have to be highly
suppressed. Extensions of the Standard Model that implement LFV
should keep such processes automatically small and allow for
new-physics particles with moderate masses.
In the quark sector, these issues can nicely be accommodated
with the Minimal Flavour Violation (MFV) hypothesis \cite{MFV}. 
Analogously to the quark sector, Minimal Lepton Flavour Violation (MLFV)
can be formulated as an effective field theory in which the lepton Yukawa couplings are the
only sources of flavour violation \cite{CGIW}. In order to additionally explain the smallness of neutrino masses with the
help of the see-saw mechanism and estabishing the requirements for leptogenesis, the MFV hypothesis in the lepton sector includes also lepton number violation at some high scale. Avoiding additional flavour violation, the three heavy right-handed Majorana neutrinos introduced are
degenerate in mass.


\section{General Picture}
Since radiative corrections spoil the degeneracy of
the Majorana masses \cite{CIP06,Branco:2006hz}, it is appropriate to combine the MLFV hypothesis with a choice
of a scale at which the Majorana masses are exactly degenerate \cite{Branco:2006hz}. A natural
choice for the degeneracy scale is the GUT scale. Then the mass splittings of the Majorana neutrinos at the Majorana scale required for resonant leptogenesis \cite{Pilaftsis:2003gt} are
induced radiatively.
This mechanism to obtain the BAU is called radiative resonant leptogenesis (RRL) \cite{GonzalezFelipe:2003fi}.

\section{CP Violation at high and low Energies}
Considering first the case with CP violation in the neutrino Yukawa
couplings present at high and low energies the situation is as follows \cite{Branco:2006hz}:
\begin{itemize}
\item The baryon asymmetry of the universe can be generated of the right order of magnitude with RRL independent
of the Majorana scale. The inclusion of flavour effects \cite{flavour,Nardi:2006fx} in the Boltzmann equations below a certain Majorana
 scale is relevant. 
\item Correlations between the generation of the BAU and LFV decays such as $\mu \to e
\gamma$ or ratios of such processes are very weak. Therefore MLFV is not as predictive as the
corresponding framework in the quark sector.
\item A flavour specific treatment allows for successful leptogenesis in the special case of
no high-energy CP violation which is in accordance with the
findings of
\cite{Nardi:2006fx,lowCP}.
\end{itemize}

\section{Special Case of exclusively low-energy CP Violation}
\begin{figure}[b!]
\includegraphics[width=7.9cm]{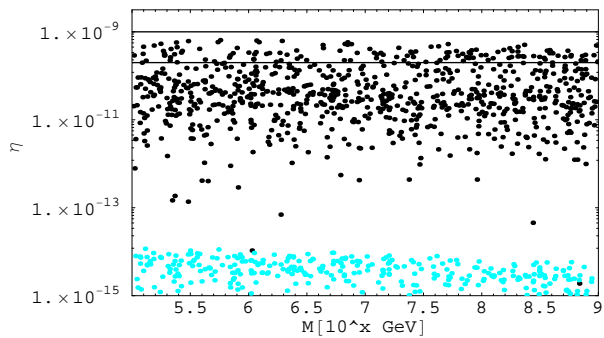}\hspace{2pc}%
\begin{minipage}[b]{14pc}\vspace{-1cm}\caption{\label{fig:BAUlow}The BAU versus the Majorana scale up to $10^9$ GeV in the case of exclusively low-energy CP violation present \cite{Uhlig:2006xf}.
  The \emph{black points} correspond to the three-flavour estimate, the \emph{light-blue points} to the single-flavour
  solution \cite{Pilaftsis:2003gt}. The two black lines mark where the BAU is of the right order
  of magnitude.}
\end{minipage}
\end{figure}
\begin{center}
\begin{figure}
\includegraphics[width=7.8cm,clip]{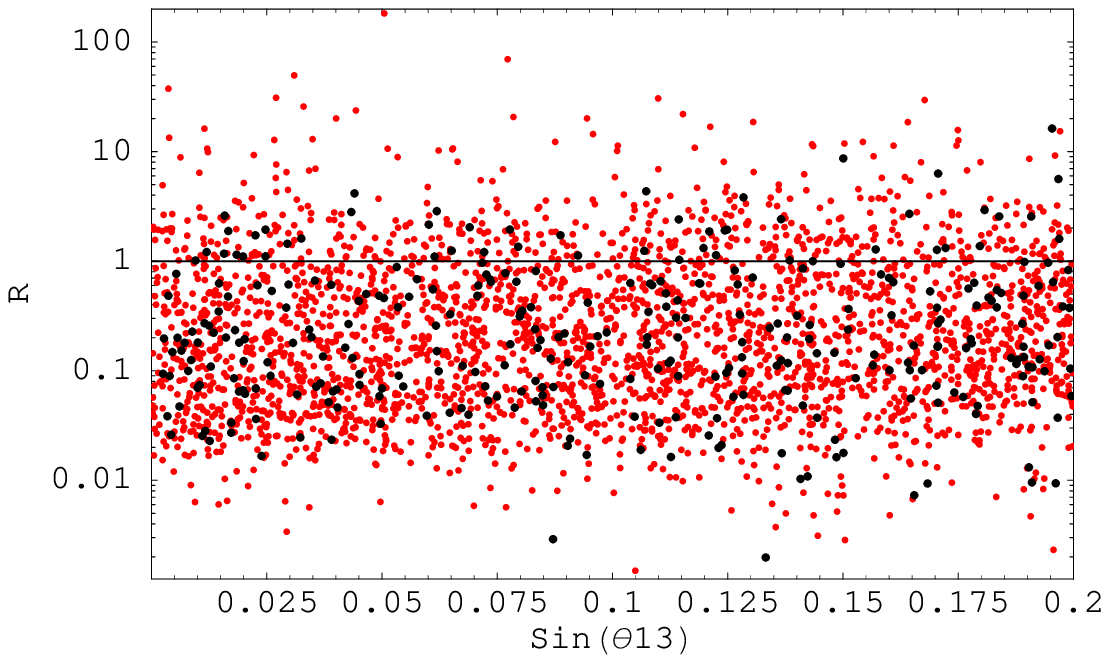}
\includegraphics[width=7.8cm,clip]{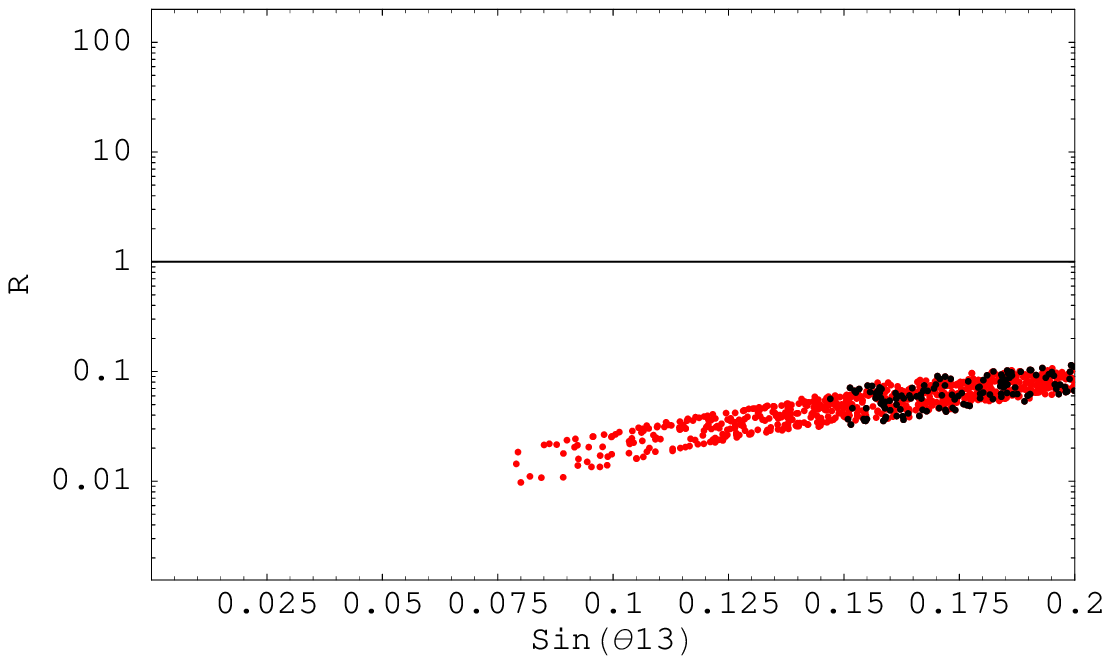}
\caption{${\rm R}=B(\mu \to e \gamma)/B(\tau \to \mu \gamma)$ versus
  $\sin{(\theta_{13})}$ for the
  general analysis \cite{Branco:2006hz} including high-energy CP
  violation (left plot) and with exclusively high-energy CP violation (right plot) \cite{Uhlig:2006xf} where
  R is clearly below 1. The \emph{black points} fulfill the
  leptogenesis constraint, the constraint on  $\sin{(\theta_{13})}$ in the lower case can be
  read off.}\label{fig:R}
\end{figure}
\end{center}

We study then the implications of a successful leptogenesis in the case of exclusively low-energy CP violation with the PMNS matrix being the only source of CP
violation (the $R$ matrix in $Y_\nu$ of the Casas-Ibarra Parametrization \cite{Casas:2001sr} is real), which can be obtained provided flavour effects are taken
into account ($M \lesssim 10^{9}-10^{12}$ GeV) \cite{Branco:2006hz}. We find that the right amount of the baryon asymmetry of
the universe can be 
generated in the flavour sensitive regime (see figure \ref{fig:BAUlow}) under the following conditions \cite{Uhlig:2006xf}:
\begin{itemize}
\item The light neutrino masses have a normal hierarchy,
\item there is at least one non-vanishing Majorana phase,
\item $\sin{(\theta_{13})}\gtrsim 0.13$,
\item $m_{\nu, {\rm lightest}}\lesssim
0.04$ eV.
\end{itemize}
If these constraints are fulfilled, we find strong correlations among
ratios of charged LFV processes. For example the ratio of $B(\mu \to e \gamma)/B(\tau \to \mu
\gamma)$ which varies over many orders of magnitude when high-energy CP violation is present,
is found then to be clearly below one (see figure \ref{fig:R}).\\
Therefore the specific case of exclusively low-energy CP violation turns out to be much more
predictive than the general one and offers constraints that are testable in low-energy experiments.
\ack{
I would like to thank G.C.Branco, A.J.Buras and M.N.Rebelo for discussions.
This research was partially supported by the German Bundesministerium f\"ur Bildung und Forschung under contract 05HT6WOA and by the Graduiertenkolleg M\"unchen.}

\smallskip
\end{document}